\newcommand\apjcls{1}
\newcommand\aastexcls{2}
\newcommand\othercls{3}

\newcommand\papercls{\aastexcls}
\documentclass[tighten,times,twocolumn]{aastex62}  

\if\papercls \apjcls
\usepackage{apjfonts}
\else\if\papercls \othercls
\usepackage{epsfig}
\usepackage{margin}
\usepackage{times}
\fi\fi
\usepackage{ifthen}
\usepackage{natbib}
\usepackage{bm}
\usepackage{amssymb, amsmath}
\usepackage{appendix}
\usepackage{etoolbox}
\usepackage[T1]{fontenc}
\usepackage{paralist}
\usepackage{newtxtext,newtxmath}
\if\papercls \apjcls
\newcommand\aas{\ref@jnl{AAS Meeting Abstracts}}
\newcommand\dps{\ref@jnl{AAS/DPS Meeting Abstracts}}
\newcommand\maps{\ref@jnl{MAPS}}
\else\if\papercls \othercls
\usepackage{astjnlabbrev-jh}
\fi\fi

\if\papercls \aastexcls
\hypersetup{citecolor=blue, 
            linkcolor=blue, 
            menucolor=blue, 
            urlcolor=blue}  
\else
\usepackage[
bookmarks=true,           
bookmarksnumbered=true,   
colorlinks=true,          
citecolor=blue,           
linkcolor=blue,           
menucolor=blue,           
urlcolor=blue,            
linkbordercolor={0 0 1},  
pdfborder={0 0 1},
frenchlinks=true]{hyperref}
\fi
\if\papercls \othercls

\else

\fi

\providecommand{\adsurl}[1]{\href{#1}{ADS}}

\makeatletter
\patchcmd{\NAT@citex}
  {\@citea\NAT@hyper@{%
     \NAT@nmfmt{\NAT@nm}%
     \hyper@natlinkbreak{\NAT@aysep\NAT@spacechar}{\@citeb\@extra@b@citeb}%
     \NAT@date}}
  {\@citea\NAT@nmfmt{\NAT@nm}%
   \NAT@aysep\NAT@spacechar\NAT@hyper@{\NAT@date}}{}{}

\patchcmd{\NAT@citex}
  {\@citea\NAT@hyper@{%
     \NAT@nmfmt{\NAT@nm}%
     \hyper@natlinkbreak{\NAT@spacechar\NAT@@open\if*#1*\else#1\NAT@spacechar\fi}%
       {\@citeb\@extra@b@citeb}%
     \NAT@date}}
  {\@citea\NAT@nmfmt{\NAT@nm}%
   \NAT@spacechar\NAT@@open\if*#1*\else#1\NAT@spacechar\fi\NAT@hyper@{\NAT@date}}
  {}{}
\makeatother

\makeatletter
\DeclareRobustCommand{\lowcase}[1]{\@lowcase#1\@nil}
\def\@lowcase#1\@nil{\if\relax#1\relax\else\MakeLowercase{#1}\fi}
\makeatother

\DeclareSymbolFont{UPM}{U}{eur}{m}{n}
\DeclareMathSymbol{\umu}{0}{UPM}{"16}
\let\oldumu=\umu
\renewcommand\umu{\ifmmode\oldumu\else\math{\oldumu}\fi}

\if\papercls \othercls

\else

\fi

\let\oldsim=\sim
\renewcommand\sim{\ifmmode\oldsim\else\math{\oldsim}\fi}
\let\oldpm=\pm
\renewcommand\pm{\ifmmode\oldpm\else\math{\oldpm}\fi}
\newcommand\by{\ifmmode\times\else\math{\times}\fi}

\newbox{\wdbox}
\renewcommand\c{\setbox\wdbox=\hbox{,}\hspace{\wd\wdbox}}
\renewcommand\i{\setbox\wdbox=\hbox{i}\hspace{\wd\wdbox}}

\newcount\timect
\newcount\hourct
\newcount\minct
\newcommand\now{\timect=\time \divide\timect by 60
         \hourct=\timect Cltiply\hourct by 60
         \minct=\time \advance\minct by -\hourct
         \number\timect:\ifnum \minct < 10 0\fi\number\minct}

\catcode`@=11

\newcommand\comment[1]{}

\newcommand\commenton{\catcode`\%=14}

\renewcommand\math[1]{$#1$}
\newcommand\mathshifton{\catcode`\$=3}

\let\atab=&
\newcommand\atabon{\catcode`\&=4}

\let\oldmsp=\sp
\let\oldmsb=\sb
\def\sp#1{\ifmmode
           \oldmsp{#1}%
         \else\strut\raise.85ex\hbox{\scriptsize #1}\fi}
\def\sb#1{\ifmmode
           \oldmsb{#1}%
         \else\strut\raise-.54ex\hbox{\scriptsize #1}\fi}
\newbox\@sp
\newbox\@sb
\def\sbp#1#2{\ifmmode%
           \oldmsb{#1}\oldmsp{#2}%
         \else
           \setbox\@sb=\hbox{\sb{#1}}%
           \setbox\@sp=\hbox{\sp{#2}}%
           \rlap{\copy\@sb}\copy\@sp
           \ifdim \wd\@sb >\wd\@sp
             \hskip -\wd\@sp \hskip \wd\@sb
           \fi
        \fi}
\def\msp#1{\ifmmode
           \oldmsp{#1}
         \else \math{\oldmsp{#1}}\fi}
\def\msb#1{\ifmmode
           \oldmsb{#1}
         \else \math{\oldmsb{#1}}\fi}

\def\supon{\catcode`\^=7}

\def\subon{\catcode`\_=8}

\def\supsubon{\supon \subon}

\newcommand\actcharon{\catcode`\~=13}

\newcommand\paramon{\catcode`\#=6}

\comment{And now to turn us totally on and off...}

\newcommand\reservedcharson{ \commenton  \mathshifton  \atabon  \supsubon 
                             \actcharon  \paramon}

\catcode`@=12
\reservedcharson

\if\papercls \apjcls

\else

\fi

\newcommand\chisq{\ifmmode{\chi\sp{2}}\else\math{\chi\sp{2}}\fi}
\newcommand\redchisq{\ifmmode{ \chi\sp{2}\sb{\rm red}}
                    \else\math{\chi\sp{2}\sb{\rm red}}\fi}
\newcommand\Teq{\ifmmode{T\sb{\rm eq}}\else$T$\sb{eq}\fi}
\newcommand\mjup{\ifmmode{M\sb{\rm Jup}}\else$M$\sb{Jup}\fi}
\newcommand\rjup{\ifmmode{R\sb{\rm Jup}}\else$R$\sb{Jup}\fi}
\newcommand\msun{\ifmmode{M\sb{\odot}}\else$M\sb{\odot}$\fi}
\newcommand\rsun{\ifmmode{R\sb{\odot}}\else$R\sb{\odot}$\fi}
\newcommand\mearth{\ifmmode{M\sb{\oplus}}\else$M\sb{\oplus}$\fi}
\newcommand\rearth{\ifmmode{R\sb{\oplus}}\else$R\sb{\oplus}$\fi}

\newcommand\Pran{\ensuremath{\mathrm{Pr}}}

\renewcommand{\bm}[1]{{\mbox{{\boldmath$#1$}}}}	

\begin{document}

\title{Accretion Rates and Thermohaline Convection in Polluted White Dwarfs}

\author{J. R. Fuentes}
\affiliation{\rm TAPIR, California Institute of Technology, Pasadena, CA 91125, USA}

\author{Matias Castro-Tapia}
\affiliation{\rm Department of Physics and Trottier Space Institute, McGill University, Montreal, QC H3A 2T8, Canada}

\author{Jim Fuller}
\affiliation{\rm TAPIR, California Institute of Technology, Pasadena, CA 91125, USA}

\author{Marcus N. King}
\affiliation{\rm TAPIR, California Institute of Technology, Pasadena, CA 91125, USA}

\begin{abstract}
    The atmospheres of polluted white dwarfs contain information about the composition of disrupted planetary material. However, interpreting the observed abundances requires a quantitative description of the mixing processes that transport accreted material through their envelopes. In particular, the accretion of metal-rich debris drives thermohaline convection in the star's radiative zone. Previous studies have modeled this form of turbulence as an enhanced diffusive process, often inferring mass accretion rates $\dot{M}_{\rm acc}$ that exceed X-ray constraints. Here we develop a time-dependent model that treats thermohaline mixing as a propagating front whose evolution is coupled self-consistently to the surface abundance of heavy elements. We perform hydrodynamical simulations which demonstrate that thermohaline mixing drives the composition gradient towards values slightly above marginal thermohaline stability. Using this condition, we derive analytic solutions for the mixed layer depth $h$ and surface abundance $X_{\rm surf}$, finding that both scale as $t^{1/2}$ at early times, transitioning to $h \propto t^{4/27}$ and $X_{\rm surf} \propto t^{10/27}$ at later times. Without gravitational settling, the thermohaline front never reaches a steady state, while including settling leads to an equilibrium surface abundance $X_{\rm surf, eq} \propto \dot{M}_{\rm acc}^{30/43}$. Applied to G 29--38, our model matches the observed heavy-element abundance for $\dot{M}_{\rm acc} \sim 1\times 10^{10}$--$2\times 10^{10}~\mathrm{g~s^{-1}}$, comparable to previous findings using stellar evolution codes. These rates exceed the X-ray estimate by about an order of magnitude, which may point to a partially suppressed X-ray luminosity from the accretion boundary layer. We also present a practical prescription for inferring accretion rates directly from the observed surface abundance and stellar parameters of other polluted white dwarfs.
\end{abstract}


\section{Introduction}
\label{sec:Introduction}

White dwarfs (WDs) are the evolutionary endpoints for the vast majority of stars and possess extremely high surface gravities, $g \sim 10^8\,\mathrm{cm\,s^{-2}}$, causing heavy elements to gravitationally settle out of their atmospheres on timescales ranging from days in warm hydrogen-atmosphere WDs to $\sim \mathrm{Myr}$ in cool helium-atmosphere WDs \citep{Koester2009}, far shorter than their cooling ages ($t_{\rm cool}\sim 100\,\mathrm{Myr}$--10 Gyr). As a result, the photospheres of isolated WDs are expected to be composed of nearly pure hydrogen or helium \citep{Schatzman1948}. Contrary to this expectation, large spectroscopic surveys have shown that approximately $\sim 25$--$50\%$ of isolated WDs have atmospheres polluted with heavy elements \citep[see, e.g.,][]{Zuckerman2003ApJ,Zuckerman2010, Koester2014}. 

\begin{figure*}
    \centering
\includegraphics[width=0.905\textwidth]{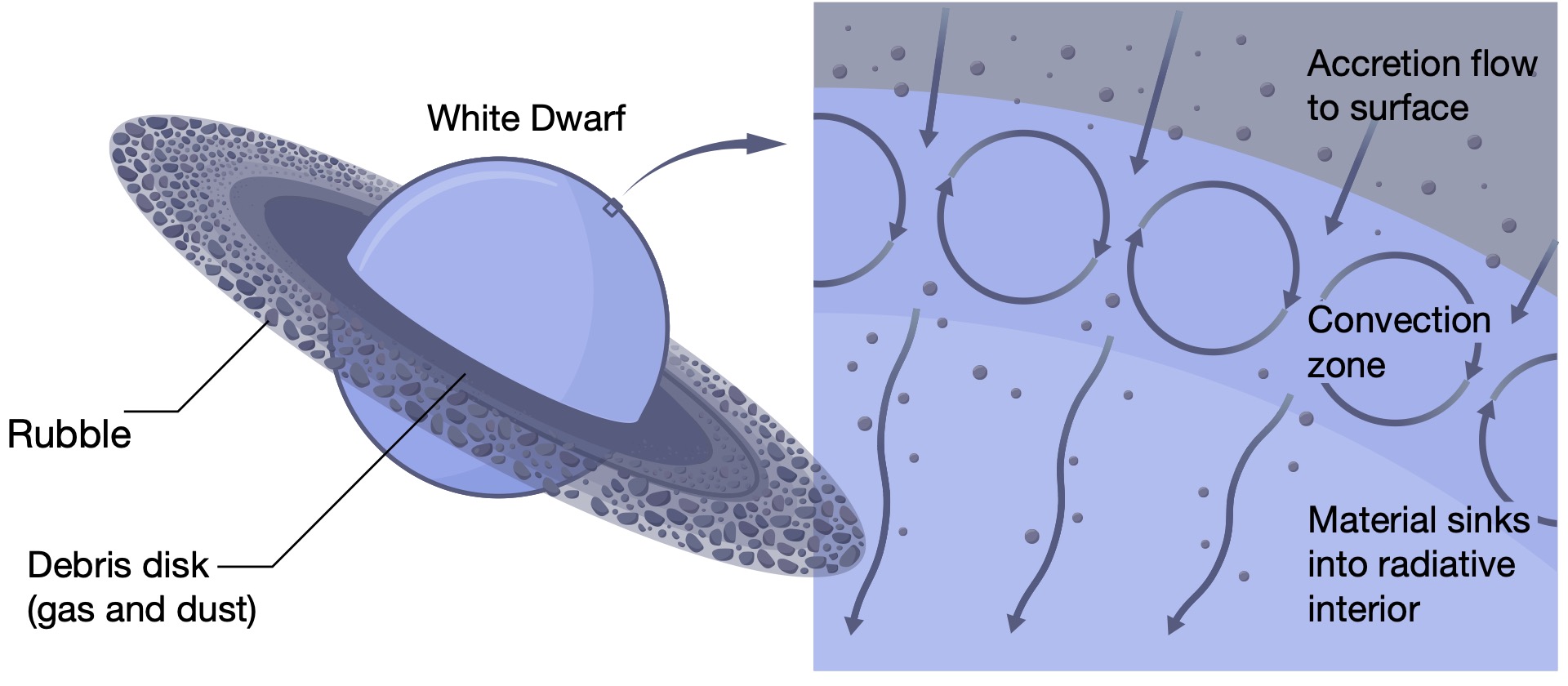}
    \caption{Cartoon illustrating the accretion of heavy elements onto the surface of a white dwarf from tidally disrupted planetesimals. Once the material reaches the star's surface, it is rapidly mixed within the surface convection zone, producing an inverse mean molecular weight gradient at its base. The accreted material subsequently sinks into the radiative interior through gravitational settling and thermohaline mixing.}
    \label{fig:placeholder}
\end{figure*}

The common interpretation is that polluted WDs are accreting debris from remnant planetary systems, typically through the tidal disruption of asteroids or minor planets that form a circumstellar disk \citep[see, e.g.,][]{Brouwers2022,Akiba2024}. This interpretation is further supported by infrared excesses associated with dusty disks, gaseous emission lines, and the inferred bulk compositions of accreted material, which resemble rocky Solar System bodies \citep[see, e.g.,][]{Farihi2011, Xu2012, Girven2012,Brinkworth2012,Madurga2024}. Polluted WDs therefore provide a unique window into the composition of exoplanetary material and the long-term evolution of planetary systems.

Inferring accretion rates and composition of these planetesimals from observed metal abundances requires an accurate understanding of the transport processes within the WD envelope. Traditionally, accretion models assume that metals are mixed only within the surface convection zone and subsequently lost by gravitational settling at its base \citep[see, e.g.,][]{Koester2009,Farihi2013,Raddi2015}. In general, the inferred accretion rates from those models range from $\dot{M}_{\mathrm{acc}} \sim 10^6$ to $10^{8}\,\mathrm{g\,s^{-1}}$. However, this picture neglects other mixing processes below the base of the surface convection zone, which would increase the
mass of the polluted region and hence the inferred accretion rate. One such process is convective overshoot, in which convective plumes penetrate into the underlying stable layer. Three-dimensional simulations of DA white dwarfs show that overshoot can extend the mixed region by about an order of magnitude in mass, raising the inferred accretion rates accordingly \citep[e.g.,][]{Cunningham2019}.

Another important process is thermohaline (fingering) convection. The accretion of high mean molecular weight material onto a lighter background produces an inverse compositional gradient that is unstable to thermohaline mixing. As shown by \cite{Deal2013} and confirmed by \cite{Wachlin2017} and \cite{Bauer2018,Bauer2019}, thermohaline mixing can transport accreted material to depths far beyond the observable surface layers. Reproducing the observed abundances then requires accretion rates several orders of magnitude larger than in models without thermohaline mixing, reaching $\dot{M}_{\mathrm{acc}} \sim 10^{13}\,\mathrm{g\,s^{-1}}$ for WDs with hydrogen-rich atmospheres \citep[see also,][]{Dwomoh2023}.

Independent constraints on the accretion rate can be obtained from X-ray observations, which probe the energy released as material accretes onto the white dwarf surface. \cite{Cunnigham2022} presented the first detection of X-ray emission from the polluted white dwarf G 29--38 using Chandra observations, later confirmed by \cite{Estrada-Dorado2023} using archival data from XMM-Newton. From the observed X-ray luminosity, they infer an instantaneous planetary material accretion rate of $\dot{M}_X \sim 10^{9}$--$3\times 10^9~\mathrm{g\,s^{-1}}$. \cite{Cunnigham2022} found that this rate is consistent with the higher accretion rates inferred from atmospheric models that include convective overshoot at the base of the surface convection zone \citep{Cunningham2019}, which extends the mixed region by roughly an order of magnitude in mass and lengthens the diffusion timescale. In contrast, accretion rates inferred from stellar evolution models including thermohaline mixing are typically an order of magnitude or more larger than the X-ray value \citep[e.g.,][]{Wachlin2017,Bauer2018}. X-ray-based accretion rates nevertheless carry their own uncertainties, because they depend on poorly constrained boundary-layer physics. For example, if the boundary layer is optically thick, the emergent X-ray luminosity may be suppressed by several orders of magnitude, as expected for cataclysmic variables \citep{NarayanPopham1993}.


Several mixing prescriptions for thermohaline convection \citep[see review by][]{Garaud2021} have been implemented in one-dimensional stellar evolution models of polluted white dwarfs to infer the composition and accretion rates of disrupted planetary material. These models treat thermohaline mixing as an effective diffusion that instantaneously mixes unstable regions without explicitly capturing the transport of heavy elements. Recent three-dimensional hydrodynamical simulations by \cite{Creswell2025}, however, demonstrate that when a constant flux of heavy elements is imposed at the top of a thermally stable layer, thermohaline convection develops as a localized front that propagates inward, progressively increasing the depth of the mixed region. 

Understanding the propagation of the thermohaline front is therefore essential for predicting the surface abundances of polluted white dwarfs. As the front deepens, the accreted material is diluted over an increasingly large mass reservoir, reducing the compositional gradient that drives the instability and slowing the advance of the front. The evolution of the mixed layer and the surface abundance are therefore intrinsically coupled. In this work, we develop a time-dependent model that follows this coupled evolution self-consistently. In Section~\ref{sec:accretion_driven_turbulence}, we use a stellar model computed with \texttt{MESA} to quantify the relevant microphysical parameters that characterize thermohaline mixing in polluted white dwarfs, providing key inputs for future hydrodynamical and magnetohydrodynamical simulations. In Section~\ref{sec:numerical}, we derive analytic scaling relations for the growth of the thermohaline front in the shallow and deep stratification limits, and assess the role of gravitational settling in establishing a long-term equilibrium. Finally, in Section~\ref{sec:conclusion}, we discuss our results in the context of previous studies, examine implications for accretion rates inferred from observations, and highlight remaining uncertainties in the accretion process.

\section{Accretion-Driven Turbulence}\label{sec:accretion_driven_turbulence}

We now estimate the microphysical properties of thermohaline convection in polluted white dwarfs. We first review the conditions under which the instability arises and identify the dimensionless parameters that govern its dynamics. We then estimate the relevant transport coefficients using an analytic envelope model calibrated against a MESA model of G 29--38. Together, these quantities are used in the front-propagation model developed in Section~\ref{sec:numerical}.

\subsection{Thermohaline instability}
\label{sec:accretion}

When metal-rich material accretes onto the WD, rapid mixing distributes this material through the surface convection zone, increasing the atmospheric abundance of heavy elements. The result is a compositionally dense layer overlying the lighter hydrogen–helium radiative interior, producing an inverted mean molecular weight gradient at the base of the convection zone. This configuration is gravitationally unstable in the compositional sense, yet, the stabilizing thermal stratification of the underlying radiative region provides a competing restoring force: thermal buoyancy resists displacement, while the compositional gradient drives it. 

The dynamics of the fluid in this region is controlled by three dimensionless parameters that describe the microscopic transport processes and the balance between stabilizing and destabilizing gradients: the Prandtl number, the Lewis number, and the stability ratio. The Prandtl number, $\mathrm{Pr} = \nu / k_T$, compares viscous to thermal diffusion. The Lewis number, $\mathrm{Le} = k_T / D$, compares thermal to atomic diffusion. The stability ratio measures the ratio of the stabilizing thermal stratification to the destabilizing compositional gradient produced by the accreted heavy elements

\begin{equation}\label{eq:R0}
R_0 = \dfrac{\delta(\nabla -\nabla_{\mathrm{ad}})}{\phi \nabla_\mu} = \dfrac{|N^2_T|}{|N^2_\mu|}\,,
\end{equation}
where
\begin{equation}
N_T^2 = -\delta\dfrac{g}{H_P}(\nabla - \nabla_{\rm ad})\,,
\qquad
N_\mu^2 = \phi\dfrac{g}{H_P}\nabla_\mu
\end{equation}
are the thermal and compositional contributions to the Brunt–Väisälä frequency, respectively. Here, $\nabla = d\ln T/d\ln P$ is the thermal gradient, $\nabla_\mu = d\ln \mu/d\ln P$ is the mean molecular weight gradient induced by accretion, and $\nabla_{\mathrm{ad}}$ is the adiabatic gradient. The thermodynamic derivatives are defined as $\delta = -(\partial \ln \rho/\partial \ln T)_{P,\mu}$ and $\phi = (\partial \ln \rho/\partial \ln \mu)_{P,T}$.

Linear stability analyses \cite[e.g.,][]{Baines1969} have shown that the fluid is unstable to overturning convection for $R_0 < 1$, whereas for $R_0 > \mathrm{Le}$ the fluid remains stable, and no instabilities arise. In the intermediate regime, $1 < R_0 < \mathrm{Le}$, the fluid becomes unstable to thermohaline convection. This is a weaker form of turbulence, often encountered in oceanography, where elongated structures known as ``salt fingers'' develop. These fingers transport high mean molecular weight material downward and low mean molecular weight material upward.  Despite being less efficient than overturning convection and unable to fully homogenize the fluid, thermohaline mixing weakens the compositional gradient, driving it toward the critical value for marginal thermohaline stability \cite[see, e.g.,][]{Fuentes2023,Castro-Tapia2024}.

\subsection{Properties of Thermohaline Mixing} \label{sec:thermohaline}
\begin{figure*}
    \centering
    \includegraphics[width=\textwidth]{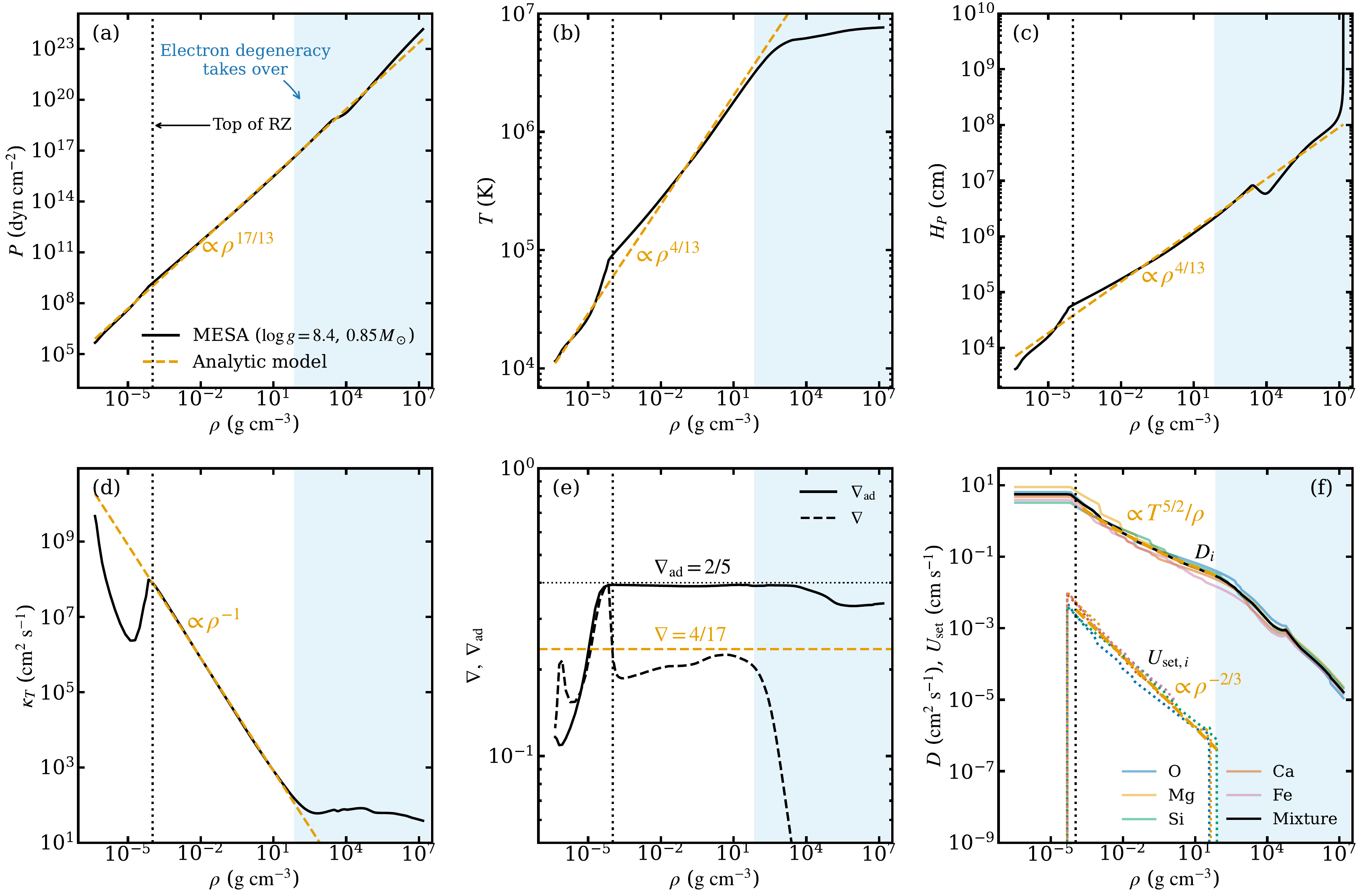}
    \caption{Interior properties of a \texttt{MESA} model of the white dwarf G 29--38, as a function of density $\rho$. As expected, the scalings from a radiative envelope based on an ideal gas equation of state break down in the degenerate core ($\rho \gtrsim 70\ \mathrm{g\,cm^{-3}}$), where electron degeneracy pressure dominates and $P \propto \rho^{5/3}$ instead. RZ stands for radiative zone. The bottom right panel shows the atomic diffusion coefficients $D_i$ (solid) and gravitational settling velocities $U_{\rm{set},i}$ (dotted) of the accreted elements from \texttt{MESA}, together with the mass-weighted diffusivity of the mixture (black). The dashed orange lines show $D\propto T^{5/2}/\rho$, as predicted by Equation (2) in \cite{Deal2026}, and the fit $U_{\rm set} = U_{\rm mix}(\rho/\rho_0)^{-2/3}$. Settling velocities are shown only above the depth reached by the accreted material after 50 kyr.}
    \label{fig:model_comparison}
\end{figure*}
To estimate the properties of thermohaline convection in polluted WDs, we first require the thermal diffusivity, viscosity, and chemical diffusion coefficient. To obtain those, we model the envelope of the star as a fully ionized, hydrogen-dominated radiative layer. At the low densities characteristic of this region, the gas is well described by an ideal equation of state, $P \propto \rho T$. Assuming that heat is transported by photon diffusion and adopting Kramers' opacity $\kappa \propto \rho T^{-7/2}$, the thermal stratification of the envelope is set by the radiative gradient $dT/dr \propto \kappa \rho/ T^3$. Solving hydrostatic balance, $dP/dr = -\rho g$, in a thin layer where the gravitational acceleration is approximately constant, yields the following scalings for the stellar structure

\begin{align}
P(\rho) \propto \rho^{17/13},\quad T(\rho)\propto \rho^{4/13},\quad k_T(\rho) \propto \rho^{-1}\,.
\end{align}
We find that these solutions agree well with a MESA model of G 29--38 in the non-degenerate radiative zone (see Figure\,\ref{fig:model_comparison})\footnote{The MESA inlists and radial profiles are publicly available at \dataset[DOI:10.5281/zenodo.23675360]{https://doi.org/10.5281/zenodo.23675360}}. Gaia-based determinations favor $\log g \approx 8.0$ \citep{McCleery2020} and $M_{\rm WD}\approx 0.6M_{\odot}$, consistent with recent asteroseismic constraints \citep[e.g.,][]{Uzundag2023}. Here, we instead adopt $M_{\rm WD} \approx 0.85M_{\odot}$ ($\log g\approx 8.4$) for consistency with the abundance analysis of \citet{Xu2014}, and with previous thermohaline models of G 29--38 \citep[e.g.,][]{Wachlin2017, Bauer2018}. The model was computed by evolving a $0.85\,M_\odot$ white dwarf from \citet{Bauer2023} to an effective temperature $T_{\rm eff}\approx 11722~\mathrm{K}$ using \texttt{MESA} v26.04.1 \citep{Paxton2011, Paxton2013, Paxton2015,
Paxton2018, Paxton2019, Jermyn2023}, along with the inlists of \citet{Dwomoh2023} to ensure sufficient resolution in the outer layers.

The resulting model has radius $R_{\rm WD} \approx R_\oplus$, surface gravity $g \approx 2.5\times 10^8~\mathrm{cm~s^{-2}}$, a hydrogen-layer mass of $M_{H}/M_{\rm WD}\approx10^{-4}$, 
and a thin surface convection zone of mass $M_{\rm CZ} \sim 10^{-14}~M_{\rm WD}$. We emphasize that $M_{\rm CZ}$ is highly uncertain for G 29--38. Published $T_{\rm eff}$ values span nearly $1000$~K, over which $M_{\rm CZ}$ varies by about three orders of magnitude, and convective overshoot may extend the mixed region further \citep{Cunningham2019}. Because thermohaline mixing carries the accreted metals tens of kilometers below the convection zone, thermohaline models are much less sensitive to this uncertainty than classical accretion--diffusion models.

Just below the convection zone, partial ionization of hydrogen keeps $\nabla\approx\nabla_{\rm ad}$, so we take as
reference the shallowest depth where hydrogen is fully ionized ($\nabla_{\rm ad} \simeq 2/5$ and $\nabla \simeq 4/17$ at $\approx 1.5$~km below the stellar surface) corresponding to the radial location $r_0 \approx 6.68 \times 10^8\,\mathrm{cm}$. At this location, the mass density, temperature, pressure scale height, and thermal diffusivity are

\begin{gather}
\nonumber \rho \approx 1
\times 10^{-4}\,\mathrm{g\,cm^{-3}}, \quad
T  \approx 9.1\times 10^{4}\,\mathrm{K}\\
H_{P}\approx 5.8\times 10^4~\mathrm{cm},\quad
k_{T}  \approx 7.9\times 10^{7}\,\mathrm{cm^{2}\,s^{-1}}.
\end{gather}

Since the kinematic viscosity is not directly available from the \texttt{MESA} output, we adopt the prescription of \citet{Spitzer1965}, appropriate for the non-degenerate outer layers where ion–ion collisions dominate the momentum transport

\begin{gather}\label{eq:Spitzer}
\nu = 2.2\times 10^{-15} \mathrm{g\,cm^{-1}\,s^{-1}\,K^{-5/2}}\dfrac{T^{5/2}}{\rho \ln\Lambda}\,,
\end{gather}
where

\begin{gather}
\ln\Lambda = 23.5 - \dfrac{1}{2}\ln n_e +  \dfrac{3}{2} \ln T\,,
\end{gather}
is the Coulomb logarithm, and $n_e\sim \rho/m_p$ is the number density of electrons. Using the reference values at $r_0$, we find $\ln\Lambda \approx 17.8$ and $\nu \approx 3 \,\rm{cm^2\,s^{-1}}$. As expected, the Prandtl number is very small, $\mathrm{Pr} =  \nu/\kappa_T \sim  4\times 10^{-8}$.  

For the atomic diffusion coefficient and gravitational settling velocities, we follow \cite{Bauer2018} and include an accretion episode of bulk-Earth material at a constant rate $\dot{M}_{\rm acc} \approx 10^9~\mathrm{g~s^{-1}}$ (accreted mass fractions $X_{\rm Fe} = 0.307,\, X_{\rm O} = 0.295,\, X_{\rm Mg} = 0.199,\, X_{\rm Si} = 0.153,\, X_{\rm Ca} = 0.046$), and take the diffusion coefficients $D_i$ and settling velocities $U_{\rm{ set},i}$ computed by \texttt{MESA}. As shown in Figure~\ref{fig:model_comparison}\,(f), the diffusivities of these elements agree with each other within a factor of two. Throughout the radiative envelope they are well described by  $D\propto T^{5/2}/\rho$, the temperature and density dependence of the prescriptions of \cite{Michaud2015} and \cite{Deal2026}  (see their Equations 4.58 and 2, respectively), with the proportionality constant depending on the ionic charge and Coulomb logarithm, whose variations with depth largely cancel. The settling velocities agree with each other as well, and decrease inward approximately as $U_{\rm set} \propto \rho^{-2/3}$. We show them only above the depth reached by the accreted material, since below it each element is effectively absent and the \texttt{MESA} velocities are not meaningful.

Since thermohaline convection is driven by the mean molecular weight gradient, rather than adopting a single representative element, we use the mass-weighted diffusivity of the accreted mixture, $D = \sum_i X_i\,D_i$. At $r_0$ we find $D \approx 4~\mathrm{cm^2~s^{-1}}$, giving a Lewis number $\mathrm{Le} = k_T/D \approx 2\times 10^7$, in agreement with the values reported by \cite{Bauer2018} (see their Figure 2). Since the individual diffusivities differ from this value by at most a factor of two, adopting a single representative element instead (e.g., oxygen, magnesium, or iron) changes $\mathrm{Le}$ by a similar factor and does not affect our conclusions.

Thermohaline mixing operates over a wide range of spatio-temporal scales, with slow, large-scale modes and fast, small-scale modes contributing differently to the composition transport \citep{Skoutnev2026}. In the limit of small Prandtl numbers and large stability ratios, such that $1 \ll R_0 \ll \mathrm{Pr}^{-1}$, the vertical transport is consistent with the classical mixing prescription commonly adopted in stellar evolution models, $D_{\rm th} \sim k_T/R_0$, \citep[e.g.,][]{Ulrich1972,Kippenhahn1980}. Typical conditions in G~29--38 suggest $R_0 \sim 100$--$1000$ \cite[see Figure~2 of][]{Bauer2018}, while $\mathrm{Pr} \sim 4\times 10^{-8}$, placing the thermohaline instability well within this regime. However, since the composition gradient evolves during accretion and after thermohaline mixing develops, $R_0$ may approach the regime $R_0 \sim \mathrm{Pr}^{-1}$, where viscous effects can modify the transport and lead to the ``Brown regime'', where $D_{\rm th}\sim k_T (\mathrm{Pr}/R_0)^{1/2}$ \citep{Brown2013,Fraser2026}. The thermohaline model that we develop below does not depend on either transport prescription.

\section{Thermohaline Front Propagation in a Stratified Stellar Envelope}\label{sec:numerical}

In this section, we present a simplified model that predicts the extent and surface abundance of a thermohaline layer in an accreting white dwarf. We adopt the stellar structure model for the radiative envelope developed in Section~\ref{sec:thermohaline}, where the density and pressure scale height as functions of depth $z$ (measured from the top of the radiative zone, with $z=0$ at $r=r_0$) are
\begin{align}\label{eq:density_profile}
\rho(z) = \rho_0\left(1+\frac{z}{z_0}\right)^{13/4}, \quad 
H_P(z) = H_0\left(1 + \frac{z}{z_0}\right),
\end{align}
with $z_0 = (17/4)k_B T_0/(m_p g) = (17/4)H_0$.

In analogy with thermal convection, where efficient transport drives the thermal gradient toward the adiabatic value, we assume that thermohaline mixing drives the composition gradient towards its critical value for marginal thermohaline stability. This assumption is approximately valid as long as thermohaline mixing occurs on timescales short enough to substantially weaken the composition gradient. More generally, however, sustaining thermohaline convection requires a composition gradient that sits above the critical value, as discussed by \cite{Fuentes2023}. Under this assumption, the composition profile is determined entirely by the marginal stability condition, together with the boundary conditions $X_Z(0) = X_{\rm surf}(t)$ and $X_{Z}(h) = 0$

\begin{gather}\label{eq:marginal_stability}
\frac{dX_Z}{dz} = \left(\frac{dX_Z}{dz}\right)_{\rm crit} = \dfrac{\nabla - \nabla_{\rm ad}}{H_P}\left(\dfrac{D}{k_T}\right), 
\end{gather}
where we used the critical stability ratio $R_{\rm crit} = k_T/D$, and that $\nabla_\mu \sim H_P dX_Z/dz$ for an ideal gas with $X_Z\ll 1$. 
Alternatively, the composition profile can be obtained from the steady-state flux balance with a turbulent diffusivity $D_{\rm th}$, as in stellar evolution codes, without assuming marginal stability. In this case, the shape of the profile depends on the adopted prescription, which could change our quantitative results by a factor of order unity. To avoid relying on a diffusion approximation of thermohaline mixing, whose efficiency remains uncertain, we use the critical gradient to obtain the composition profile.

The layer depth is obtained from mass conservation. For steady accretion at rate $\dot{M}_{\rm acc}$, and assuming negligible transport of heavy elements across the lower boundary, the total heavy-element mass in the layer evolves as
\begin{gather}\label{eq:mass_conservation}
\frac{d}{dt}\left(4\pi r_0^2 \int_0^{h(t)} \rho(z)\, X_Z(z,t)\, dz\right) = \dot{M}_{\rm acc}~.
\end{gather}
In reality, molecular diffusion and gravitational settling transport heavy elements through the lower boundary of the thermohaline layer, introducing a sink term into Equation~\eqref{eq:mass_conservation}. Within the mixed region, however, thermohaline transport is much faster than gravitational settling, so the downward leakage remains small compared with the accretion-driven flux that sustains the front. We therefore neglect this term when deriving the analytic solutions and assess its importance later in Section~\ref{sec:equilibrium}.

Together, the marginal-stability condition, the composition profile it implies, and mass conservation form a closed system that determines both the surface abundance $X_{\rm surf}(t)$ and the layer depth $h(t)$. In the following, we derive analytical scalings in the limits of shallow and deep layers, finding $h \propto t^{1/2}$ and $X_{\rm surf}\propto t^{1/2}$ at early times, transitioning to $h \propto t^{4/27}$ and $X_{\rm surf}\propto t^{10/27}$ as the front penetrates into the stratified envelope.

\subsection{Shallow Layer Limit (Constant Density)}\label{sec:shallow_layer}

We first consider the limit of a shallow thermohaline layer, $h/z_0 \ll 1$, for which the density and pressure scale height remain approximately constant throughout the mixed region. In this limit, $\rho(z) \approx \rho_0$, and all thermodynamic quantities can be approximated by their values at the top of the radiative zone, $z=0$.

In the shallow layer limit, the critical composition gradient is spatially uniform, set by the stabilizing thermal stratification and the microscopic diffusivities at $z = 0$

\begin{gather}\label{eq:critical_gradient_shallow}
\left(\dfrac{dX_Z}{dz}\right)_{\rm crit,0} = \dfrac{\nabla_{\mu,\rm crit,0}}{H_0} = \left(\dfrac{\nabla - \nabla_{\rm ad}}{H_0}\right)\left(\dfrac{D_0}{\kappa_{T,0}}\right) < 0.
\end{gather}
Integrating Equation~\ref{eq:critical_gradient_shallow} gives a linear composition profile,
\begin{equation}\label{eq:X_linear}
 X_Z(z) = X_{\rm surf}(t)\left(1 - \dfrac{z}{h}\right),
\end{equation}
with the surface abundance and layer depth related by
\begin{equation}\label{eq:Xsurf_h_shallow}
X_{\rm surf}(t) =  \dfrac{|\nabla_{\mu,\rm crit,0}|}{H_0}\,h.
\end{equation}
Substituting into mass conservation (Equation~\ref{eq:mass_conservation}) then gives

\begin{gather}
h(t) = \left(\dfrac{2H_0}{\left|\nabla_{\mu,\rm crit,0}\right|}\dfrac{F_{\rm acc}\, t}{\rho_0}\right)^{1/2},\\
X_{\rm surf}(t) = \left(\dfrac{2\left|\nabla_{\mu,\rm crit,0}\right|}{H_0}\dfrac{F_{\rm acc}\, t}{\rho_0}\right)^{1/2},
\end{gather}
where $F_{\rm acc} = \dot{M}_{\rm acc}/4\pi r^2_0$ is the accretion flux at the top of the front. While these solutions exhibit a $t^{1/2}$ scaling that resembles a diffusive growth, the underlying dynamics are not diffusive in nature. In particular, the evolution is not governed by a constant effective diffusivity and therefore does not correspond to the classical scaling $h(t)\sim (D_{\rm th} t)^{1/2}$. We verified these shallow-layer scalings with three-dimensional hydrodynamical simulations in the Boussinesq approximation, which recovers $h\propto t^{1/2}$ and $X_{\rm surf}\propto t^{1/2}$ and a nearly linear composition profile within the mixed layer (see Appendix~\ref{sec:appendix_1}--\ref{sec:appendix_2}).

\subsection{Deep Layer Limit (Stratified Polytrope)}
\begin{figure*}
\centering    \includegraphics[width=\textwidth]{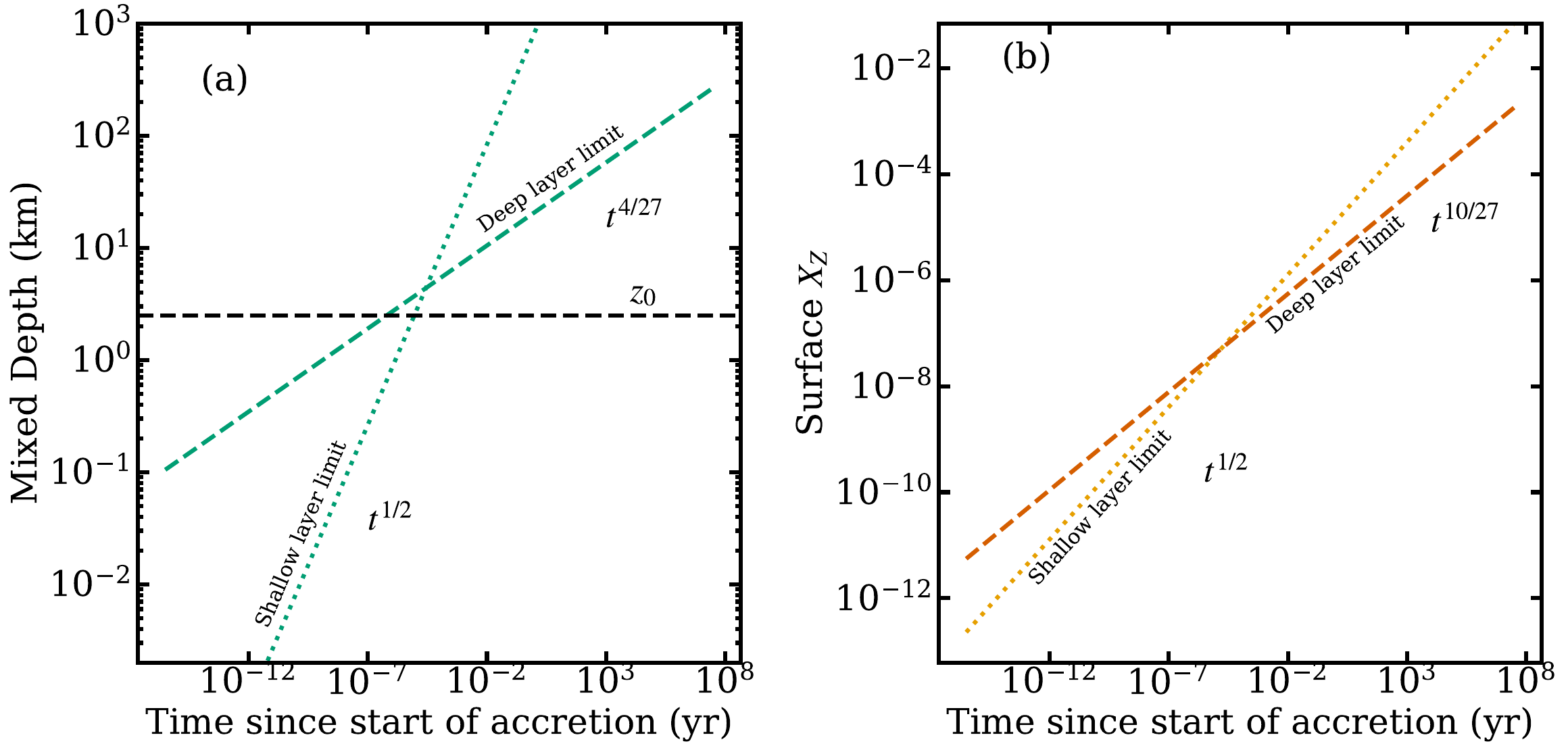}
    \caption{Evolution of the thermohaline layer using the parameters of G 29--38 for an accretion rate $\dot{M}_\mathrm{acc} = 10^{10}\, \mathrm{g\,s^{-1}}$. Panel (a): Thermohaline layer depth $h(t)$ as a function of time since the onset of accretion. Panel (b): Mass fraction of heavy elements at the surface, $X_\mathrm{surf}(t)$ over the same period. In both panels,  the dotted and dashed lines show the shallow-layer and deep-layer asymptotic scalings, respectively. The horizontal dashed line in panel (a) marks the scale height $z_0 = (17/4)H_0 \approx 2.5\ \mathrm{km}$, which sets the transition between the two regimes.}
    \label{fig:analytic_solution}
\end{figure*}
We now consider the opposite limit of a deep thermohaline layer, $h/z_0 \gg 1$, where the density varies strongly with depth, $\rho(z)\approx \rho_0(z/z_0)^{13/4}$, and the material properties are well approximated by power-law functions of density (see Figure~\ref{fig:model_comparison}).

The critical composition gradient resembles that of the shallow layer case (Equation~\ref{eq:critical_gradient_shallow}), with the addition of a density-dependent factor arising from the depth dependence of the pressure scale height, as well as the thermal and chemical diffusivities (see Figure~\ref{fig:model_comparison})

\begin{align}
\nonumber\left(\dfrac{dX_Z}{dz}\right)_{\rm crit} &= \left(\dfrac{\nabla - \nabla_{\rm ad}}{H_0}\right)\left(\dfrac{D_0}{\kappa_{T,0}}\right)\left(\dfrac{\rho}{\rho_0}\right)^{6/13},\\
&= \left(\dfrac{\nabla_{\mu,\rm crit,0}}{H_0}\right)\left(\dfrac{z}{z_0}\right)^{3/2}. \label{eq:marginal_profile_deep}
\end{align}
Note that the critical gradient steepens with depth, meaning progressively more accreted material is required to advance the front into denser layers. Integrating Equation~\eqref{eq:marginal_profile_deep} gives

\begin{equation}\label{eq:X_deep}
X_Z(z) = X_{\rm surf}(t)\left[1 - \left(\dfrac{z}{h}\right)^{5/2}\right],
\end{equation}
with the surface abundance and layer depth related by
\begin{equation}\label{eq:Xsurf_h_deep}
    X_{\rm surf} = \dfrac{2}{5}|\nabla_{\mu,\rm crit,0}|\left(\dfrac{z_0}{H_0}
    \right)\left(\dfrac{h}{z_0}\right)^{5/2}.
\end{equation}
Substituting into mass conservation (Equation~\ref{eq:mass_conservation}) then gives
\begin{gather}
h(t) = C_1 H_0\left(\dfrac{F_{\rm acc}\,t}{\rho_0 H_0 
\left|\nabla_{\mu,\rm crit,0}\right|}\right)^{4/27}, 
\label{eq:h_deep}\\
X_{\rm surf}(t) = C_2\left|\nabla_{\mu,\rm crit,0}\right|
\left(\dfrac{F_{\rm acc}\,t}{\rho_0 H_0 \left|\nabla_{\mu,\rm crit,0}\right|}\right)^{10/27},
\label{eq:Xsurf_deep}
\end{gather}
where $C_1 \approx 4.55 $ and $C_2 \approx 2$.

We emphasize that unlike the shallow-layer limit, where the Boussinesq approximation permits direct numerical validation, the deep stratified regime is beyond the reach of current numerical simulations. The thermohaline front spans many density scale heights and evolves on timescales much longer than the local mixing time. The scalings presented here are therefore physically motivated but remain to be tested by future simulations.

Figure~\ref{fig:analytic_solution} shows the analytical solutions for the thermohaline layer, using the stellar parameters of G 29--38 derived in Section~\ref{sec:thermohaline} and an accretion rate of $\dot{M}_\mathrm{acc} = 10^{10}\, \mathrm{g\ s^{-1}}$. Except for a brief initial phase lasting $\ll 1$ yr after the onset of accretion, the evolution of the thermohaline front proceeds entirely in the deep stratified regime. 

We emphasize that direct comparison with observations is challenging because the analytical predictions above do not reach a steady state, as both the thermohaline layer depth and the surface abundance continue to increase with time. This occurs because the model does not include any mechanism for the removal of heavy elements from the bottom of the thermohaline layer, allowing metals to continuously accumulate within the layer. In realistic white dwarf envelopes, gravitational settling provides such a sink by transporting metals out of the thermohaline region, potentially establishing an equilibrium 
between accretion and downward diffusion. In what follows, we incorporate gravitational settling into the model and use the resulting equilibrium solutions to compare our predictions with observations.

\subsection{Equilibrium including Gravitational Settling}\label{sec:equilibrium}

The downward flux of heavy elements due to gravitational settling is

\begin{align}
F_{\rm set} = \rho U_{\rm set} X_Z,
\end{align}
where $U_{\rm set}$ is the gravitational settling velocity. In equilibrium, this flux balances the flux of accreted material, providing a steady-state surface abundance.

As discussed in Section~\ref{sec:thermohaline} and shown in Figure~\ref{fig:model_comparison}\,(f), $U_{\rm set}$ is well described by

\begin{align}
U_{\rm set}(\rho) = U_0\left(\dfrac{\rho}{\rho_0}\right)^{-2/3},
\end{align}
where $U_0 \approx 3\times10^{-3}~\mathrm{cm~s^{-1}}$ is the mass-weighted settling velocity of the accreted mixture at the top of the thermohaline layer in our MESA model of G 29--38. Since the settling velocity is determined by the interaction between the sinking heavy elements and the surrounding hydrogen plasma, $U_0$ is independent of the accretion rate.

We expect thermohaline mixing to dominate over most of the layer, with gravitational settling taking over near its bottom. In order to reach an equilibrium, the composition profile at large depths (where gravitational settling dominates) must yield a flux that is constant with depth and equal to the accretion flux, i.e., $F_{\rm set} = F_{\rm acc}$ which entails

\begin{align}
    F_{\rm acc} &= \rho U_{\rm set} X_{\rm set} \nonumber \\
    &= \rho_0 U_0 \left(\dfrac{z}{z_0}\right)^{13/12} X_{\rm set} \, .
\end{align}
This implies that the composition profile in the settling-dominated region is

\begin{equation}
\label{eq:xset}
    X_{\rm set} = \frac{F_{\rm acc}}{\rho_0 U_0} \bigg(\frac{z}{z_0}\bigg)^{-13/12} \, .
\end{equation}

The transition between the thermohaline and settling-dominated regions can be found by matching $X_{\rm set}$ from Equation \eqref{eq:xset} and the deep thermohaline composition profile given by Equation~\eqref{eq:X_deep}, which we denote $X_{\rm th}$. Additionally, the derivatives $dX_{\rm set}/dz$ and $dX_{\rm th}/dz$ must also be equal at the transition point. Combining these requirements and doing some algebra yields the transition depth $z_c$,

\begin{equation}
    \frac{z_c}{h} = \bigg(\frac{13}{43}\bigg)^{2/5} \simeq 0.62 \, .
\end{equation}

\begin{figure}
    \centering
    \includegraphics[width=\columnwidth]{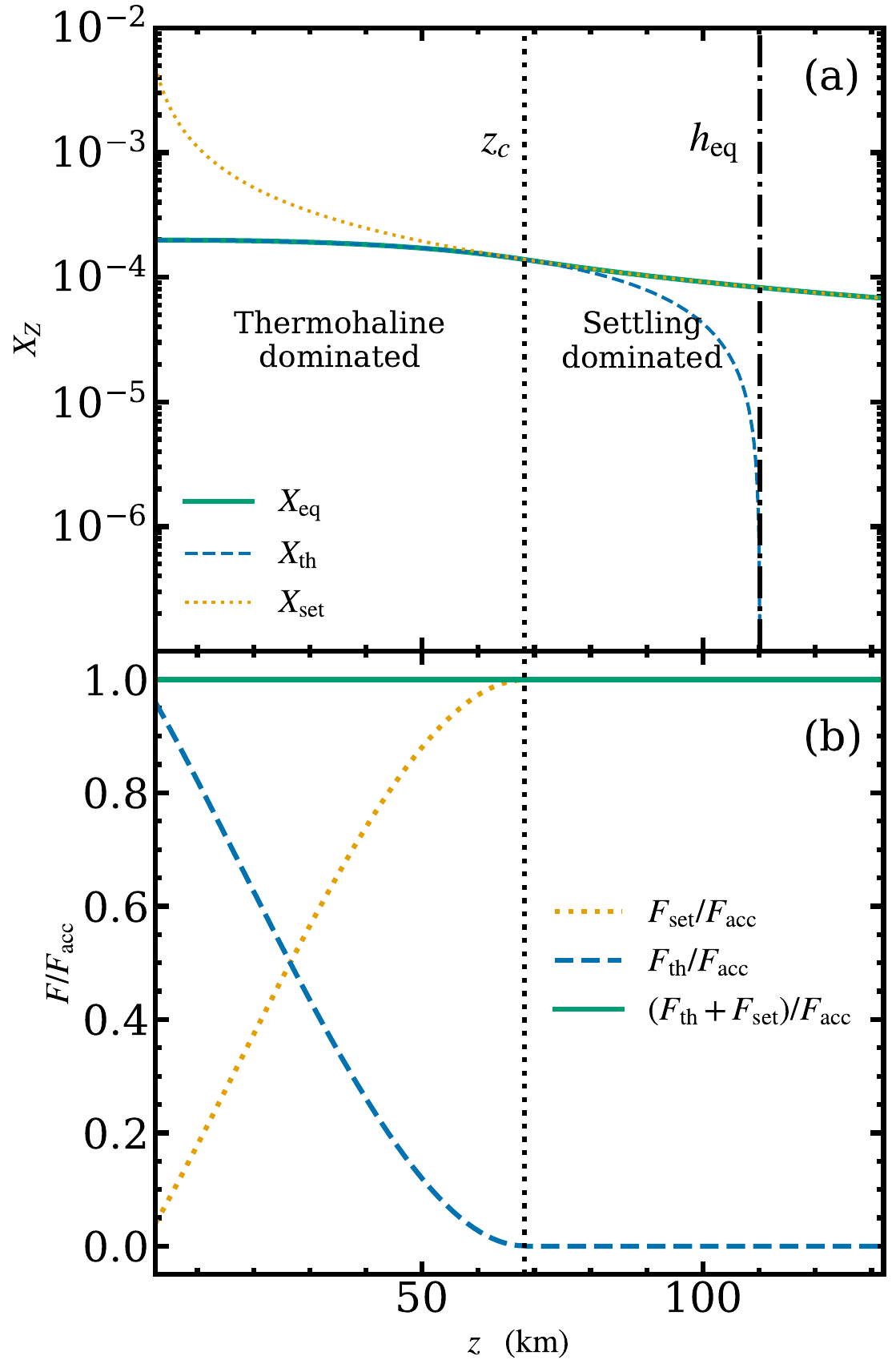}
    \caption{Equilibrium profiles as a function of depth $z$. Panel (a): Heavy-element mass fraction $X_Z$. Dotted and dashed curves show the profiles expected from gravitational settling alone ($X_{\rm set}$) and thermohaline convection alone ($X_{\rm th}$). The solid curve is the equilibrium profile $X_{\rm eq}$ obtained when both processes operate together and the resulting total flux becomes constant (see text). The dotted vertical line marks the critical depth $z_c$ below which settling dominates over thermohaline mixing, and the dot-dashed line marks the equilibrium mixed-layer depth $h_{\rm eq}$. Panel (b): Corresponding heavy-element fluxes at equilibrium, normalized to the accretion flux $F_{\rm acc}$. Dashed and dotted curves show the thermohaline ($F_{\rm th}$) and settling ($F_{\rm set}$) contributions, whereas the solid curve shows the total flux.}
    \label{fig:profiles}
\end{figure}
Above the transition depth, $z/h \lesssim 0.62$, the downward flux is dominated by thermohaline mixing, but with an increasing contribution from gravitational settling as the depth approaches $z_c$. Below the transition depth, $z/h \gtrsim 0.62$, the composition gradient is too small to induce thermohaline mixing, and the downward flux is entirely due to gravitational settling. Figure~\ref{fig:profiles} shows the expected equilibrium vertical profiles of $X_Z$ and the heavy-element fluxes, together with the individual contributions from thermohaline convection and gravitational settling.

We can then determine the equilibrium values of the surface composition and thermohaline depth by demanding that the settling flux at the transition depth carry the accretion flux $F_{\rm acc}$. Since the two composition profiles match at $z_c$, the settling flux can equivalently be evaluated using the thermohaline composition profile from Equation~\eqref{eq:X_deep} at the transition. This gives,

\begin{equation}
F_{\rm acc} = \rho_0 U_0 X_{\rm surf,eq} \left(\dfrac{z_c}{z_0}\right)^{13/12}\left[1-\left(\dfrac{z_c}{h}\right)^{5/2}\right].
\end{equation}
Combining this condition with Equation \eqref{eq:Xsurf_h_deep} and using $z_c/h \simeq 0.62$, we obtain

\begin{gather}
X_{\rm surf,eq} \approx 2.16|\nabla_{\mu,\rm crit,0}|\left(\dfrac{F_{\rm acc}}{\rho_0 U_0 |\nabla_{\mu,\rm crit,0}|}\right)^{30/43},\label{eq:X_eq}\\
    h_{\rm eq} \approx 4.7H_0\left(\dfrac{F_{\rm acc}}{\rho_0 U_0 |\nabla_{\mu,\rm crit,0}|}\right)^{12/43}.\label{eq:h_eq}
\end{gather}
Before settling becomes important, the layer follows the deep-layer solution (Equations~\ref{eq:h_deep}--\ref{eq:Xsurf_deep}). A natural timescale to reach equilibrium is therefore the time at which Equation~\eqref{eq:Xsurf_deep} reaches $X_{\rm surf,eq}$, which gives

\begin{align}\label{eq:t_eq}
t_{\rm eq} \approx 1.2 \left(\dfrac{H_0}{U_0}\right)\left(\dfrac{F_{\rm acc}}{\rho_0 U_0 |\nabla_{\mu,\rm crit,0}|}\right)^{38/43} \approx \dfrac{z_c}{U_{\rm set}(z_c)}~.
\end{align}
This is the gravitational settling time evaluated at the transition depth, as expected since in equilibrium  the heavy elements in the thermohaline layer are removed by settling at $z_c$.

\begin{figure}
    \centering
    \includegraphics[width=\columnwidth]{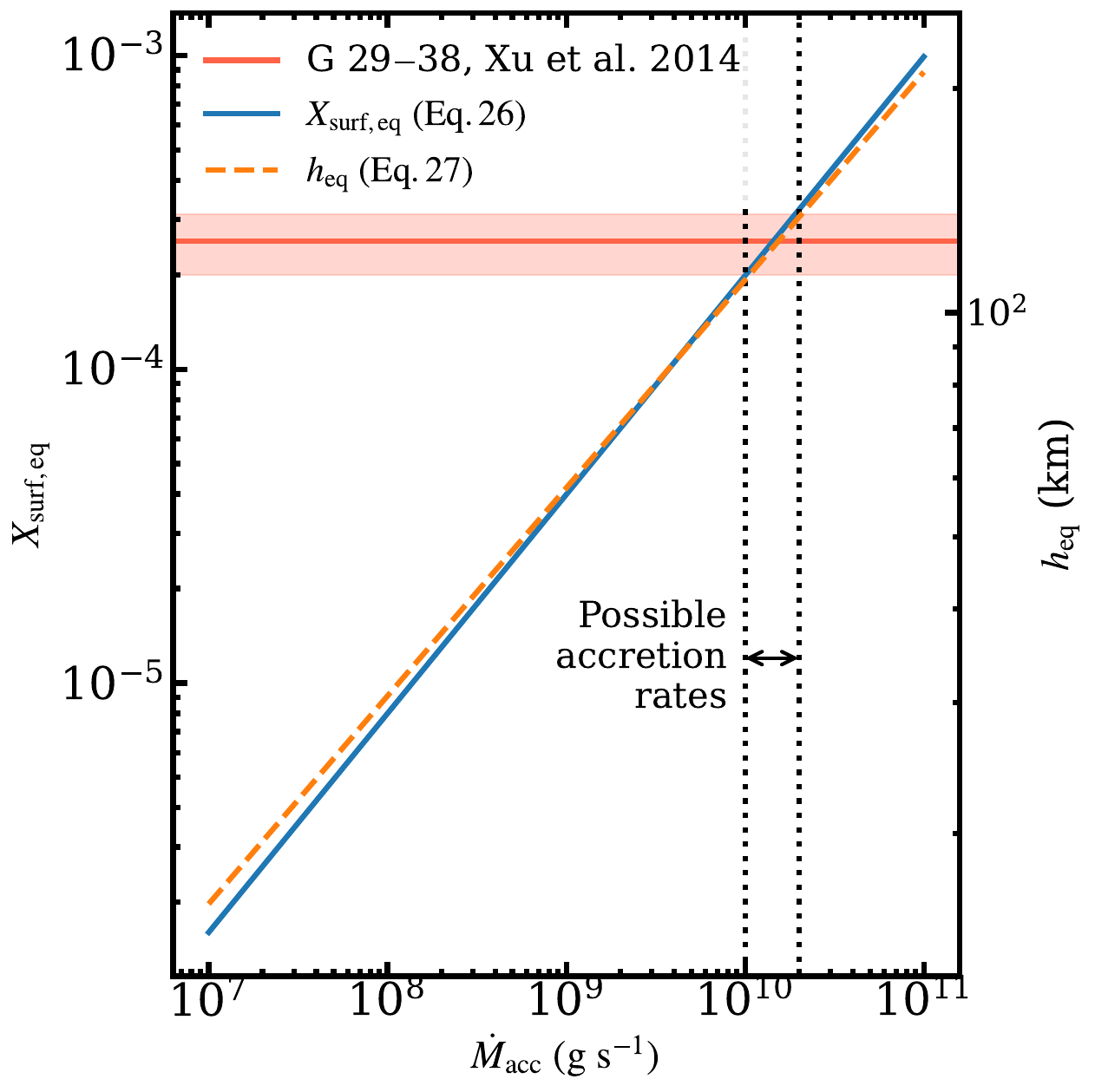}
    \caption{Predicted equilibrium mixed depth and surface abundance (Equation~\ref{eq:X_eq}--\ref{eq:h_eq}) as a function of the accretion rate. The red horizontal band shows the observed total heavy-element mass fraction in G~29--38 from \cite{Xu2014}, with the width reflecting the $1\sigma$ uncertainty propagated from the individual element abundances.}
\label{fig:Xeq}
\end{figure}

Plugging in values for G 29--38 (those in Figure~\ref{fig:model_comparison} evaluated at $r_0$, the top of the thermohaline region), our predictions suggest that accretion rates $\dot{M}_{\rm acc} \sim 1\times 10^{10}$--$2\times 10^{10}~\mathrm{g~s^{-1}}$ are required to reproduce the observed heavy-element mass fraction $X_{Z,\rm obs} \approx 2.56 \times 10^{-4}$ \citep{Xu2014} (see Figure~\ref{fig:Xeq}), provided accretion has been sustained for $\approx 10^{5}$ yr (Equation~\ref{eq:t_eq}). Evaluating in an alternative \texttt{MESA} model of G29-38 that matches $T_{\rm{eff}}\approx11770$ K, $\log{g}\approx8.0$, and $M_{\rm{WD}}\approx0.6M_{\odot}$ \citep[to account for some of the values reported for G29-38 through the years; see, e.g, the Montreal White Dwarf Database;][]{Dufour2017}, gives $\dot{M}^{\rm inf}_{\rm acc}\approx1.8\times{10^{10}}\ \rm{g\ s^{-1}}$, which is consistent with our $\log{g}\approx8.4$ model. These accretion rates are comparable to those found in previous studies of G 29--38 treating thermohaline convection as an enhanced diffusive process using stellar evolution codes \citep{Wachlin2017,Bauer2018,Buchan2025}. They exceed the X-ray estimate of \citet{Cunnigham2022} by a factor of 3--20 (see Section~\ref{sec:conclusion}).

We also find that over the range of accretion rates for which the predicted surface abundance is consistent with observations, the thermohaline region extents to a depth $h\sim 110$--$130$ km, reaching densities $\rho\sim 5$--$25~\mathrm{g~cm^{-3}}$. This remains below the density at which electron degeneracy becomes important ($\rho\sim 70~\mathrm{g~cm^{-3}}$, see Figure~\ref{fig:model_comparison}), so the ideal gas approximation is still reasonable over the region where the accreted material is mixed. Additionally, the layer is substantially deeper than $z_0 \approx 2.5$ km, such that the deep layer limit applies well.

Finally, the equilibrium solution also provides a convenient prescription for inferring accretion rates directly from the observed surface abundance. Inverting Equation~\eqref{eq:X_eq} and using $F_{\rm acc} = \dot{M}_{\rm acc}/4\pi R_{\rm WD}^2$ gives

\begin{align}\label{Mdot_formula}
\nonumber \dot{M}^{\rm inf}_{\rm acc}
&\approx 10^{9}~\mathrm{g~s^{-1}} \left(\dfrac{X_{\rm surf}}{10^{-4}}\right)^{43/30}\left(\dfrac{R_{\rm WD}}{R_{\rm \oplus}}\right)^2\left(\dfrac{\rho_0}{ 10^{-4}~\mathrm{g~cm^{-3}}}\right)\\
&\times \left(\dfrac{U_0}{10^{-3}~\mathrm{cm~s^{-1}}}\right)\left(\dfrac{|\nabla_{\mu,\rm crit,0}|}{10^{-8}}\right)^{-13/30}.
\end{align}
where $\nabla_{\mu,\rm crit,0} = (\nabla-\nabla_{\rm ad})\,(D_0/k_{T,0})$, and we have used reference values that should be re-evaluated for each individual white dwarf.  This relation provides a simple way to estimate the accretion rate for other polluted white dwarfs once their surface abundance and the corresponding stellar parameters at the top of the thermohaline region are specified.

\section{Discussion} \label{sec:conclusion}

\subsection{Summary of our results}

We have developed a time-dependent model of thermohaline mixing in the outer layers of polluted white dwarfs and applied it to conditions appropriate for G 29--38. Our model couples the growth of the thermohaline layer depth $h(t)$ to the evolution of the surface heavy-element abundance $X_{\rm surf}(t)$ through mass conservation and a marginal stability condition throughout the whole thermohaline region. We derive analytic scaling relations in the shallow ($h \ll z_0$) and deep ($h \gg z_0$) stratification limits, finding $h \propto t^{1/2}$ and $X_{\rm surf} \propto t^{1/2}$ at early times, transitioning to $h \propto t^{4/27}$ and $X_{\rm surf} \propto t^{10/27}$ as the front deepens into the stratified envelope. Neither the layer depth nor the surface abundance reaches a true steady state in the absence of gravitational settling.

When gravitational settling is included, however, an equilibrium is established between the accretion flux and the settling-driven removal of heavy elements from the thermohaline layer. In this case, the equilibrium surface abundance scales as $X_{\rm surf, eq} \propto \dot{M}_{\rm acc}^{30/43}$. For conditions appropriate to G 29--38, our model requires $\dot{M}_{\rm acc} \sim 1\times 10^{10}$--$2\times 10^{10}~\mathrm{g~s^{-1}}$ to reproduce the observed heavy-element abundance. These values are comparable to the accretion rates inferred by \cite{Wachlin2017} and \cite{Bauer2018} using stellar evolution models, but are a factor of 3--20 larger than the X-ray constraint of $1\times 10^9$--$3\times 10^9~
\mathrm{g~s^{-1}}$ from \cite{Cunnigham2022}. Lower accretion rates predict equilibrium abundances below the observed value.

We emphasize that our model is independent of the choice of the thermohaline mixing prescription (Brown or Kippenhahn), suggesting that the long-term evolution is not sensitive to the specific parameterization of the finger velocities.

\subsection{Limitations and future work}

Several aspects of the model could be improved in future work. One important limitation is the assumed form of the composition profile within the thermohaline 
layer. In this study, we derived the profile by assuming that the composition gradient is approximately the critical gradient for marginal stability. While this 
assumption is supported by our numerical simulations in the shallow-layer limit, its validity in the deep stratified regime--where density varies strongly with 
depth--remains to be tested. Direct numerical simulations of thermohaline convection in a strongly stratified medium would be needed to validate the predicted composition profiles and the associated scaling relations.

A related limitation is our treatment of gravitational settling. We construct the equilibrium composition profile by assuming that thermohaline mixing dominates in the upper part of the layer, while gravitational settling dominates at greater depths. This piecewise construction provides an analytic estimate of the equilibrium surface abundance and mixed depth, but does not explicitly solve the coupled transport problem in which both processes operate simultaneously throughout the layer. In particular, the structure of the transition region and the extent to which settling contributes to the flux above the transition may differ from our analytic estimation. Our treatment also neglects atomic diffusion, whose timescale relative to gravitational settling is $\tau_{\rm diff}/\tau_{\rm set} \sim U_{\rm set} L/D \sim 100$, using $L \sim  40~\mathrm{km}$ (the extent of the region below $z_c$), $\rho\sim 10~\mathrm{g~cm^{-3}}$, $U_{\rm set} \sim 10^{-6}~\mathrm{cm~s^{-1}}$, and $D \sim 0.05~\mathrm{cm^2~s^{-1}}$ (values near the transition depth). Since $\tau_{\rm diff}/\tau_{\rm set} \gg 1$, we expect the role of atomic diffusion to be subdominant. A more accurate solution treating settling, diffusion, and thermohaline mixing self-consistently may change the numerical coefficients in Equations \eqref{eq:X_eq} and \eqref{eq:h_eq} by order unity, but we do not expect it to alter the power law scaling of our results.

Since our study focuses on constraining accretion rates through mass conservation, we treat the mass fraction of heavy elements as a single evolving variable. This is appropriate because thermohaline convection is driven by the mean molecular weight gradient and mixes all species together. In our \texttt{MESA} model accreting a bulk-Earth composition, the settling velocities of O, Mg, Si, Ca and Fe agree to within a factor of $\approx 2$, so we adopt a single effective settling velocity for the mixture. Our model therefore does not follow the individual settling of each pollutant, which is needed to infer the relative abundances of the accreted body from the observed ones \citep[e.g.,][]{Koester2009, Koester2014}. We will extend the model to derive relative abundances together with accretion rates for a sample of polluted white dwarfs in a companion paper (Castro-Tapia et al., in preparation).

It is also possible that additional mixing processes, such as efficient compositional convection may operate in polluted white dwarfs, enhancing the compositional transport on shorter timescales. One motivation for considering such enhanced mixing is that polluted white dwarfs may undergo short episodes of intense accretion, with accretion rates of order $\sim 10^{15}\ \rm{g\ s^{-1}}$ or larger, as suggested by \citet{Farihi2012}. In this context, the accretion rates inferred from the X-ray emission of G~29--38 may represent lower limits, since the structure and efficiency of the boundary layer in polluted white dwarfs remain uncertain. In analogous accreting systems, such as cataclysmic variables and young stellar objects, optically thick boundary layers associated with high accretion rates efficiently reprocess the accretion luminosity, strongly suppressing hard X-ray emission while shifting the emergent spectrum toward softer X-ray and UV bands \citep{NarayanPopham1993,PophamNarayan1995,Wheatley2003,Suleimanov2014}.

Our results highlight the time-dependent nature of thermohaline mixing as a key factor for interpreting heavy-element abundances in polluted white dwarfs. The observed surface abundances encode the full accretion history of the system, complicating efforts to reconstruct the composition and delivery of disrupted planetary material. Nonetheless, our results suggest an equilibrium can be reached, involving the action of both thermohaline mixing and gravitational settling. As three-dimensional simulations of thermohaline convection in stratified fluids become feasible, the framework developed here will provide a physically motivated foundation for interpreting the growing census of polluted white dwarfs.

\begin{acknowledgements}
We thank the referee for a thorough and insightful report that helped improve the paper. We also thank Valentin Skoutnev and Adrian Fraser for interesting conversations on the diffusion approximation of thermohaline convection, and Evan Bauer for his advice on MESA models of G 29--38. J.R.F. is supported by the Sherman Fairchild Postdoctoral (Burke) Fellowship and the Presidential Fellowship at Caltech, as well as NASA Solar System Workings grant 80NSSC24K0927. M.C.-T. is supported by the Fonds de recherche du Québec - Nature et technologies through a doctoral scholarship (\href{	https://doi.org/10.69777/366094}{DOI:10.69777/366094}) and is a member of the Centre de Recherche en Astrophysique du Québec (AstroQuébec). M.K. acknowledges the support of the Edward C. Stone SURF Fellow donors.
\end{acknowledgements}

\appendix

\begin{figure*}
    \centering
    \includegraphics[width=\textwidth]{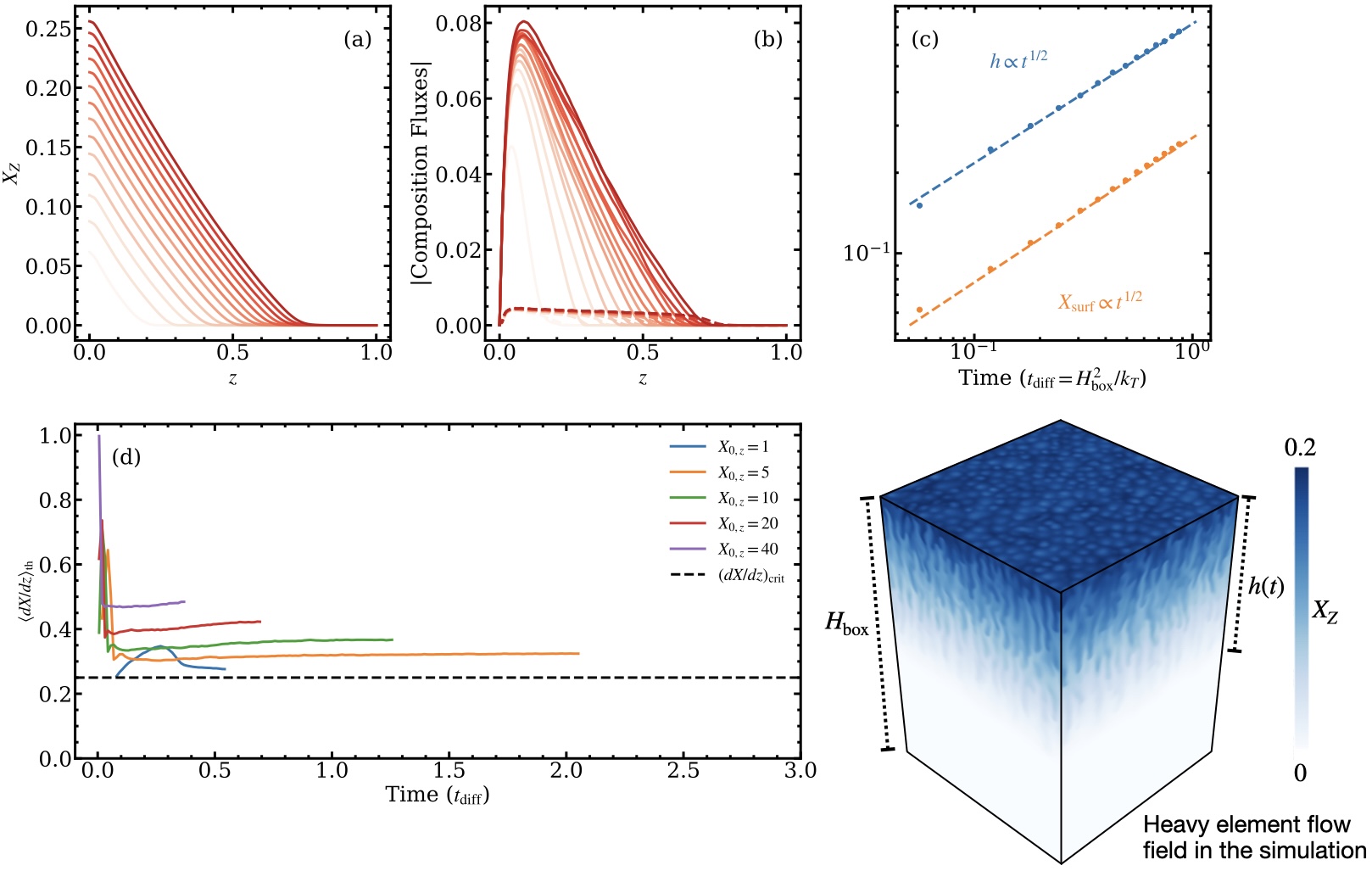}
    \caption{Results of the 3D simulations of an accretion-driven thermohaline front. Panels (a)--(c) and the 3D visualization correspond to the reference run with $X_{0,z} = 10$. Panel (a): Horizontally averaged mass fraction of heavy elements $X_Z$ as a function of depth $z$ (measured from the top of the box), at different times (light to dark red). Panel (b): Turbulent (solid) and molecular-diffusion (dashed) composition fluxes at the same times. Panel (c): Thermohaline layer depth $h$ (blue, in units of $H_{\rm box}$) and surface abundance $X_{\rm surf}$ (orange, in units of $\Delta X$) as a function of time (in units of $t_{\rm diff}$). Dashed lines are power-law fits, both consistent with $t^{1/2}$, as predicted by the shallow-layer model. Panel (d): Composition gradient averaged over the thernohaline region, as a function of time for the five values of the imposed flux. The dashed line marks the critical gradient $(dX/dz)_{\rm crit} = R_0/\mathrm{Le} = 0.25$. Bottom right: 3D visualization of the heavy elements in the simulation box, showing the thermohaline fingers carrying heavy elements downward from the source at the top.}
    \label{fig:simulation}
\end{figure*}

To validate the shallow-layer analytical model presented in Section~\ref{sec:shallow_layer}, we perform 3D simulations of thermohaline convection using the Dedalus pseudospectral framework \citep{Burns2020}. The simulation solve the incompressible fluid equations in a Cartesian domain with a localized source of heavy elements near the upper boundary that mimics continuous accretion. The simulation is intended to test the predicted evolution of an accretion-driven thermohaline front under the assumptions of the shallow-layer model, rather than to reproduce the full stratified structure of a white dwarf envelope.

\section{Fluid equations and numerical methods} \label{sec:appendix_1}

We express the fluid quantities as the sum of a linear hydrostatic background (denoted by the subscript 0) and a dynamic perturbation to the background (denoted by primes), e.g., the total temperature and composition are expressed as $T = T_0(z) + T'$, and $X = X_0(z) + X'$, respectively. The density perturbations satisfy $\rho'/\rho_0 \ll 1$, and are related to $T'$ and $X'$ through $\rho' = \rho_0(\beta X' - \alpha T')$, as demanded by the Boussinesq approximation \citep{Spiegel_Veronis_1960}. Here, $\beta$ and $\alpha$ are the coefficients of compositional and thermal contraction/expansion (both assumed positive constants), respectively.
Before presenting the fluid equations, we non-dimensionalize them using $[T] =\Delta T$, $[X] = \Delta X$ as units of temperature and composition. We use the domain's depth $H_{\rm box}$ as the unit of length, and the thermal diffusion time $\tau_T = H_{\rm box}^2/k_T$ (where $k_T$ is the thermal diffusivity) as the unit of time. By this choice, a unit of pressure corresponds to $[P] = \rho_0 (k_T/H_{\rm box})^2$. The dimensionless equations are 

\begin{gather}
\nabla \cdot \bm{u} = 0\, ,\\
\dfrac{\partial \bm{u}}{\partial t} + \bm{u}\cdot \nabla \bm{u} = - \nabla P' + \mathrm{Ra}\mathrm{Pr}\left(R^{-1}_0 X' - T'\right)\bm{\hat{z}} + \mathrm{Pr} \nabla^2\bm{u} \, ,\\
\dfrac{\partial X}{\partial t} + \bm{u}\cdot \nabla X = \mathrm{Le}^{-1} \nabla^2 X + \mathcal{S}\, ,\\
\dfrac{\partial T}{\partial t} + \bm{u} \cdot \nabla T = \nabla^2 T\, ,
\end{gather}
where $\bm{u}$ is the velocity field, and $\mathcal{S}$ is a compositional source function that mimics accretion at the upper boundary, which takes the form

\begin{equation}
\mathcal{S}(z) = \mathrm{Le}^{-1}\frac{X_{0,z}}{\delta} \exp \left( -\frac{1-z}{\delta} \right)\,,
\end{equation}
where $\delta$ specifies the thickness of the source layer and $X_{0,z}$ determines the imposed compositional flux. We adopt $\delta = 0.025$ and $X_{0,z} = {1,\, 5,\,10,\,20,\,40}$.

There are 4 dimensionless numbers that characterize the evolution of the flow. These are the Rayleigh number Ra, stability ratio $R_0$, Prandtl number Pr, and Lewis number Le, which are defined respectively as
\begin{align}
    &\mathrm{Ra} = \frac{\alpha g \Delta T H^3_{\rm box}}{\kappa_T \nu}, \quad  \!\!\!\!R_0 = \frac{\alpha \Delta T}{\beta \Delta X}, \quad
    \!\!\!\!\Pran = \frac{\nu}{k_T},\quad \!\!\!\!
    \mathrm{Le} = \frac{k_T}{D},
\end{align}
where $\nu$ is the kinematic viscosity, and $k_T$ and $D$ are the thermal and compositional diffusivities, respectively. $R_0$ is the ratio of the stabilizing and destabilizing effect of the thermal and compositional buoyancy.

We initialize the fluid with a linearly increasing distribution of temperature and zero composition,  $T_0(z) = z, \quad X_0(z) = 0$. We fix the heat flux across the box through the boundary conditions $\partial_z T|_{z=0,1} = 1$, while the composition boundary conditions are $X|_{z=0} = 0$ and $\partial_z X|_{z=1} = 0$. The boundary conditions for the velocity are impenetrable and stress free at both boundaries ($\hat{z}\cdot\bm{u} = \hat{x}\cdot\partial_z\bm{u} = \hat{y}\cdot\partial_z\bm{u} = 0$ at $z = 0,1$). 

The simulations presented here use $\mathrm{Ra} = 3\times 10^{9}$, $\mathrm{Pr} = 0.1$, and $\mathrm{Le} = 100$. Thermohaline instabilities are expected to occur when the stability ratio lies within a finite range $R_0 \in [1,~\mathrm{Le}]$ \citep[e.g.,][]{Garaud2021}, therefore we adopt $R_0 = 25$. The values of the dimensionless numbers of the simulation are less extreme than their astrophysical values in white dwarf envelopes (see Section~\ref{sec:thermohaline}). This choice is purely due to computational limitations, since resolving the true parameter regime remains prohibitively expensive.

We time-evolve equations A1--A4 using the Dedalus pseudospectral solver \citep{Burns2020} version 3, using timestepper SBDF2 \citep{wang_ruuth_2008} and CFL safety factor 0.2. All variables are represented using a Chebyshev series with 384 terms for $z \in [0, 1]$ and Fourier series with 384 terms in the periodic $x$ and $y$ directions, where $x,~y \in [0, 0.75]$. We use the 3/2-dealiasing rule in all directions, so that nonlinearities are calculated in physical space on a $576^3$ grid.
To start our simulations, we add random-noise  perturbations to the composition field sampled from a normal distribution with a magnitude of $10^{-5}$.

\section{Results}\label{sec:appendix_2}
The panels shown in Figure~\ref{fig:simulation} are obtained directly from these simulations, after transforming the coordinates from $z$ to $1-z$, to directly compare with the analytical predictions derived in Section~\ref{sec:shallow_layer}. Panels (a)--(c) show the reference run with $X_{0,z}=10$. As heavy elements are deposited near the top of the box, thermohaline convection mixes them downward and the front advances into the stable fluid (panel a). Within the layer, the composition decreases nearly linearly with depth, as predicted by the shallow-layer model (Equation~\ref{eq:X_linear}), with only a thin diffusive tail below the front. The composition flux (panel b) is carried almost entirely by the thermohaline motions, which exceed the molecular diffusion flux by more than an order of magnitude over most of the layer, and it drops nearly zero at the front. Both the layer depth and the surface abundance grow as $t^{1/2}$ (panel c).

To test whether thermohaline convection drives the layer towards the critical gradient, panel (d) shows the composition gradient averaged over the thermohaline unstable region, for five runs spanning a factor of 40 in the imposed flux ($X_{0,z} =1$--40). After an initial transient, $\langle dX/dz \rangle_{\rm th}$ settles to a nearly constant value of $\approx 0.3$--0.5, i.e., 1.2--2 times the critical gradient $(dX/dz)_{\rm crit} = R_0/\rm{Le} = 0.25$, and increases only weakly with the magnitude of the composition source (accretion rate). Thermohaline convection therefore keeps the layer close to marginal stability and a modest excess over the critical gradient is enough for the turbulent flux to carry the imposed accretion flux. We conclude that the accretion rate controls how fast the front advances rather than the gradient within the layer.

\cite{Creswell2025} conducted similar 3D Boussinesq simulations of a thermohaline front driven by a constant composition flux imposed at the upper boundary, with $\mathrm{Pr}=0.05$--0.1, $\mathrm{Le}=20$--40, and an initially unstable layer with $R_0 = 1.95$. As in our runs, the composition flux in the bulk of the layer is carried by thermohaline turbulence rather than molecular diffusion, and decreases towards the bottom of the mixed region. A notable difference is that in \cite{Creswell2025}, both the surface abundance of heavy elements and composition gradient in the mixed region decrease with time, whereas in our runs the surface abundance of metals increase with time and the gradient remains nearly constant with time. We attribute this mainly to the initial conditions. Their simulations start from a pre-existing layer whose gradient is more than 10 times the critical value, so the decreasing gradient reflects the relaxation of this strongly supercritical state. The authors did not comment on the evolution of the mixed layer depth, but from their Figures 2 and 3, $h(t)$ evolves with a power law that is much weaker than $t^{1/2}$.

\bibliography{references}{}
\bibliographystyle{aasjournal}

\end{document}